\documentclass[aps,prl,twocolumn,showpacs,amsmath,amssymb,floatfix,superscriptaddress,
]{revtex4-1}
\usepackage{graphicx}
\usepackage{soul,color}

\graphicspath{ {Figure/} } 

\begin{document}
\title{Complex active optical networks as a new laser concept}
\date{\today}

\author{Stefano Lepri}
\affiliation{Consiglio Nazionale delle Ricerche, Istituto dei Sistemi Complessi, Via Madonna del Piano 10 I-50019 Sesto Fiorentino, Italy.} 
\author{Cosimo Trono }
\affiliation{Consiglio Nazionale delle Ricerche, Istituto di Fisica Applicata "Nello Carrara", Via Madonna del Piano 10 I-50019 Sesto Fiorentino, Italy.} 
\author{Giovanni Giacomelli}
\affiliation{Consiglio Nazionale delle Ricerche, Istituto dei Sistemi Complessi, Via Madonna del Piano 10 I-50019 Sesto Fiorentino, Italy.} 

\begin{abstract} Complex optical networks containing one or more gain sections are investigated and the evidence of lasing action is reported; the emission spectrum reflects the topological disorder induced by the connections. A theoretical description well compares with the measurements, mapping the networks to directed graphs and showing the analogies with the problem of quantum chaos on graphs. We show that the interplay of chaotic diffusion and amplification leads to an emission statistics with characteristic heavy-tails: for different topologies, an unprecedented experimental demonstration of L\'evy statistics expected for random lasers is here provided for a continuous-wave pumped system. This result is also supported by a Monte-Carlo simulation based on ray random walk on the graph.
\end{abstract}

\maketitle

Pursuing laboratory investigations of complex dynamics on networks is a topic of growing interest motivated e.g. by the study of power grids and their failures \cite{Rohden2012}, the role of topology on synchronization \cite{belykh2005synchronization} or other nonlinear effects \cite{Gnutzmann2011,perakis2014small}. In this work, we present a novel scheme: the \textit{lasing network} (LANER). It consists of a complex, active optical network, whose connectivity induces a form of topological disorder and can display laser action. Besides it intrinsic interest, the system may be regarded as an optical implementation of a dynamical system on a graph \cite{Porter}. The great majority of lasers shares the same structure with a single gain section in a simple linear or ring cavity, supporting regular sets of optical modes \cite{milonni2010laser}. A somehow opposite case is the random laser, where the propagation of rays in a disordered gain medium leads to light amplification \cite{Cao2003,Wiersma2008}. The LANER generalizes to strongly connected, multiple gain setups and could also be considered as a \textit{discrete} random laser with a controllable complexity.  The robustness and flexibility of the apparatus permit to explore different configurations by an easy re-arrangement of the components; the stability of the setup allows for detailed statistical analysis as well.

\begin{figure}
\includegraphics[width=1.0\linewidth]{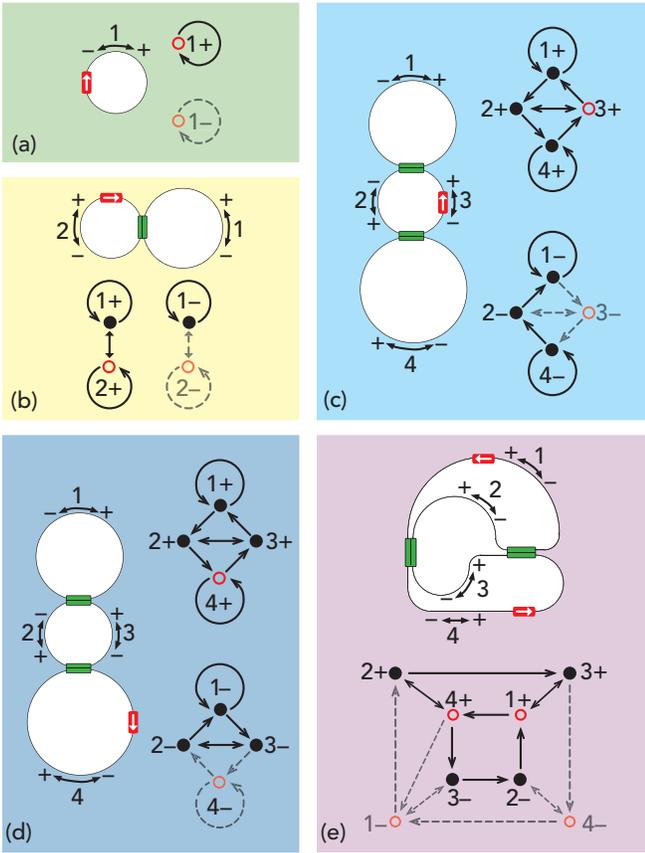}
\caption{(Color online) Simpler LANERs (white) and their equivalent graphs. Arrowed (red) segments represent oriented gain sections, numbered (black) arrows the modes in the links. Graphs: (red) circumferences denote active links modes.}
\label{fig1b}
\end{figure}

The network we consider is built with $N_s$ couplers connected by $N_l$ single-mode fibers, some of them ({\it active}) capable to provide optical amplification via stimulated emission. The couplers are standard, optical power splitters or other commercially available components such as circulators, etc. Isolators can also be used to select a propagation direction: in the present work, we will deal only with directed active fibers and $2\times2$ (four-ports) lossless couplers with no open ports, so that $N_l=2N_s$. Schemes of the simplest topologies with $N_s=1$ and $2$ realized 
experimentally are depicted in Figs.\ref{fig1b} (curves with white background).
Alongside, their representation as equivalent \textit{directed graphs} is 
given (see below).

\begin{figure*}
\includegraphics[width=0.88\linewidth,clip]{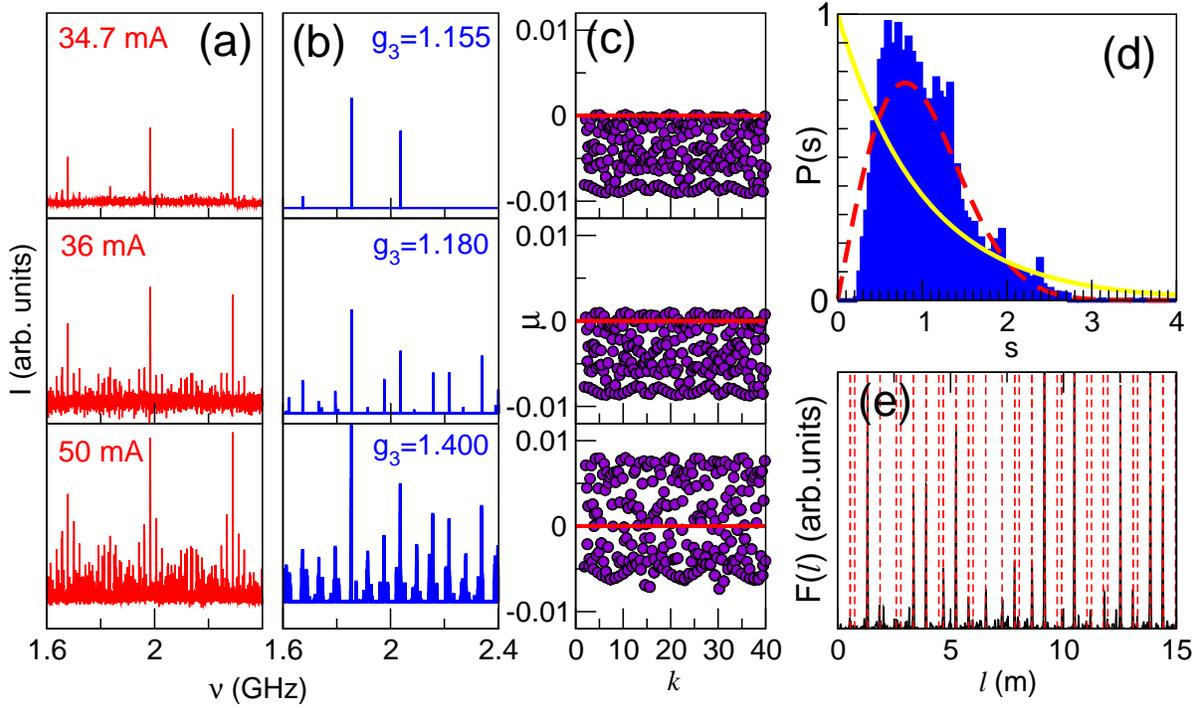}
\caption{(Color online) Results for the LANER in Fig. \ref{fig1b}c, with $L_1=9.16m,L_2=18.12m,L_3=5.24m,L_4=10.0m$. (a) Experimental intensity spectrum increasing the pump current. Numerics: (b) beatings 
for increasing $g_3$ at fixed $g_1=g_2=g_4=0.9$; (c) poles in the complex plane; (d) level spacing distribution, averaged over an ensemble of 30 graphs ($10^4$ poles each) with the same topology, $g_1=g_2=g_4=0.9,g_3=1.4$ and uniformly-distributed random $L_j$ with average $10$. Wigner (dashed) and Poisson (solid) distributions are plotted for comparison. (e) Experimental length spectrum (see Eq.(4) in the Suppl. Material) at $J=80mA$; the vertical lines represent the combinations $mL_1+n(L_2+L_3)+lL_4$ with $m,l,n$ integers.}
\label{fig2}
\end{figure*}

\begin{figure*}
\includegraphics[width=0.94\linewidth,clip]{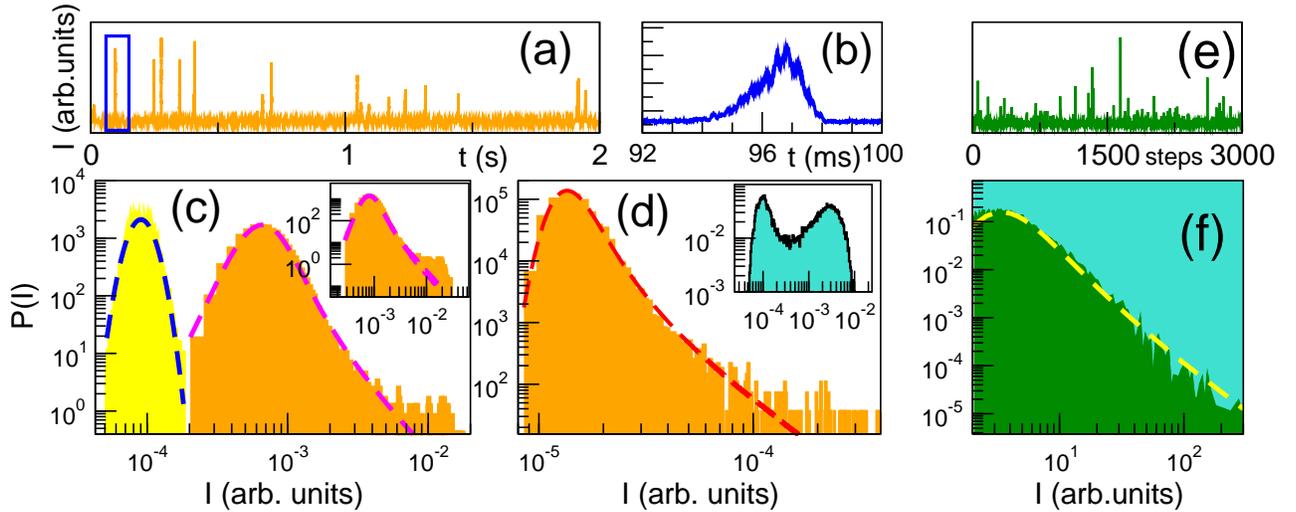}
\caption{(Color online) LANER emission dynamics and statistics.
(a)-(d): Experiment. Setup of Fig.\ref{fig1b}b, (a) time-resolved intensity at $\nu=163.57$~MHz; (b) boxed event in (a); (c) distributions of intensity for $J=42.2$ mA: left, $\nu=191.94$~MHz, (blue) dashed line, log-normal fit; right, $\nu=163.57$~MHz, (magenta) dashed line, L\'evy stable fit with $\alpha=1.4$; inset, setup of Fig.\ref{fig1b}c, distribution at $\nu=327.13$~MHz for $J=37.0$~mA, (magenta) dashed line, L\'evy stable fit with $\alpha=1.4$; 
(d) setup of Fig.\ref{fig1b}e: distribution of the intensity of the mode {\bf 3+} at $\nu=380.25$~MHz for $J_1=10$~mA (link with losses) and $J_4=55$~mA (link with gain); (red) dashed line, L\'evy stable fit with $\alpha=1.32$; inset, bistable distribution at $\nu=380.34$~MHz for $J_1=40$~mA and $J_4=60$~mA (both links have gain). Monte-Carlo simulation, geometry of Fig.\ref{fig1b}c (see Suppl. Material): (e) time series; (f) distribution of (e), (yellow) dashed line, L\'evy stable fit with $\alpha=0.995$.
}
\label{fig3}
\end{figure*}

For the theoretical description, we denote by $L_j$ the lengths of the $N_l$
fiber segments and by $g_j$ the respective gain ($g_j>1$) or loss ($g_j<1$) factors.
The observables are the envelopes of the longitudinal optical field propagating in opposite directions along each fiber.
Within a linear description, where nonlinear dispersion, gain 
saturation etc. are disregarded, the optical spectrum is determined through the $2N_l\times 2N_l$ network matrix $N=PS$ in terms of the {\it propagation} matrix $P$ (which contains the metric and topological informations) and the wave splitting at the couplers through the unitary {\it scattering} matrix $S$ (see Suppl. Material). Assuming that the field through the link $j$ is multiplied by a factor $G_j 
= g_j e^{i K L_j}$, the allowed complex wave-numbers $K$ (poles) are determined by 
the condition
$\det\left(N(K)-I\right) = 1$,
where $I$ is the $2N_l\times 2N_l$ identity matrix. 
This generalizes the usual mode-matching and threshold conditions for a laser, in the case of linear gains and no cross phase-gain coupling. Its analytic solution of is usually unfeasible even for the 
simplest networks but it can be solved numerically.  

A representation of the LANER in terms of graphs can be introduced as well, by setting all the nonzero elements of the matrix $N$ to 1 and defining the result as the adjacency matrix of the equivalent graph. In such a picture, each vertex represents a component of the field in a link, and a directed bound the linear dependence of the target on the source mode. An empty (red) vertex indicates that the related mode link is active, otherwise passive if filled (see Fig.\ref{fig1b}).  The inclusion of isolators in the active links leads to the removal of the blocked fields and therefore the corresponding vertices are removed from the equivalent graph, which is {\it pruned}. It is worth noting that the system can lase only if at least one active vertex is present in the pruned graph. At the simplest level of description, the photon dynamics in the LANER can be visualized as a Markovian random walk \cite{Burioni2005} on such graphs. 

We remark the close analogies of this scheme with the \textit{quantum graphs} which have been thoroughly investigated in the realm of quantum chaos  \cite{kuchment2004quantum,kottos1999periodic}. Indeed, in the Hamiltonian case ($g_j=1$) $N$ (termed the vertex scattering matrix) is unitary. The equation for the poles is formally equal to the one to determine the quantum spectrum of a particle moving freely along the bonds and scattered at the graph vertices. In this context, open graphs have been also considered \cite{barra2001transport,kottos2003quantum}. 
Experimentally, quantum graphs can be simulated by microwave networks \cite{hul2004experimental}. Besides the different physical
nature of our system, an important novel and controllable element 
that we deal here is the optical gain (possibly in multiple links) which allow to achieve the lasing action and investigate entirely new effects. 

We start our investigation in the setup corresponding to the network/graph of Fig.\ref{fig1b}c (Fig.1 of the Suppl. Material).  The first feature that we demonstrate is the existence of a well defined threshold for lasing, which manifests experimentally as a sudden switching of an increasingly large number of distinct peaks in the emission spectrum upon increasing of the pump current. Fig.\ref{fig2}-a shows the onset of the laser action in the experiment, while Figs.\ref{fig2}b-c report the related results from the theory. In the experiment, at low pump values the spectrum is flat and no emission is visible at any frequency; then some peaks appear and, further increasing the pump the peaks structure becomes more and more complicated. Their interaction between such increasing number of active modes is not apparently strong enough to determine a standard chaotic dynamics.
Fig.\ref{fig2}b reports the histogram of the beatings obtained from the computation of the poles $K_n=k_n-i\mu_n$: all the possible differences in frequency are evaluated for the lasing modes (those with $\mu_n>0$)  and their distribution is plotted. The result gives a qualitative estimation of the beating spectrum and directly compares to the experiment, showing the same complex hierarchical structure.
The poles $K_n$ for the same geometry are reported in the complex plane for increasing gain in the active fiber (Fig. \ref{fig2}c). Above a critical value, a set of resonances cross the real axis and the associated modes grow with rate $\mu_n>0$ and start to lase. Since we assumed an infinite gain bandwidth, the dynamics above threshold will be very high-dimensional.
\citep{kottos1999periodic}. In the experiment, the number of active modes will be limited by the gain bandwidth. For the case at hand ({\it Er$^{3+}$}-doped medium), this is estimated to be of the order of some tens of {\it nm}, so that more that $10^5$ modes may be excited. The peaks frequency is not significantly affected by the pumping, indicating that the nonlinearity only saturates the amplitude and gain dispersion is negligible.

The above results are reminiscent of highly-multimode lasers behavior; however, the possible choice of complex network topologies and arbitrary numbers of gain sections permit a generalization of the geometries for e.g. longitudinal lasers. Moreover, this sets the peculiar statistical features of our system that we discuss below.

The classical dynamics of particles on graphs is a chaotic type of diffusive process \cite{barra2001classical} that can be described theoretically by simple one-dimensional piecewise chaotic maps \cite{pakonski2001classical}. It is well known that this reflects 
in the statistical properties of the spectra as thoroughly investigated in the quantum chaos literature. For comparison, we numerically computed the nearest-neighbor level spacing distribution $P(s)$ where $s=(k_{n+1}-k_n)/\bar s$ are normalized to the average $\bar s$. All the computed $k_n$ were included in the analysis regardless of the sign of $\mu_n$ and the distributions averaged over a set of graphs with the same topology and random lengths. Even for the case above threshold,  Fig. \ref{fig2}d shows that $P(s)$ 
resembles the Wigner-Dyson distribution $\pi s\exp(-\pi s^2/4)/2$ typical of chaotic systems \cite{kottos1999periodic}. Actually, individual realizations typically 
display a level repulsion similar to what demonstrated for the Hamiltonian case \citep{barra2000level} but deviations are expected in view of the smallness the graph. 

In  Fig.\ref{fig2}e we also report the Fourier transform of the beating spectrum
$F(l)$ which is closely related to the {\it length spectrum} (see Suppl. Material); it is experimentally accessible and relates the spectra to the underlying classical periodic orbits \cite{hul2004experimental,kottos1999periodic}.
In particular, the experimental results (black peaks) well fit the theoretical prediction (vertical red lines) expressing the lengths as combination of the three {\it ring} paths of lengths $L_1$, $L_2+L_3$ and $L_4$: $L_2$ and $L_3$ are not contributing independently as expected.

The interplay of diffusion and amplification of photons in scattering media, as it occurs in random lasers \cite{Cao2003,Wiersma2008}, can lead to heavy-tailed distributions of emission intensities, characterized by  L\'evy -stable statistics \cite{uchaikin1999}. This theoretical prediction \cite{Lepri2007} 
(see also \cite{Lepri2013,raposo2015analytical}) has been confirmed in several experiments \citep{Ignesti2013,Uppu2014,gomes2016observation}.
To the extent in which the photons in our networks can be treated as
particle undergoing chaotic diffusion, it is thus plausible that 
the same phenomenology may also occur in the LANER. We consider first the configuration in Fig.\ref{fig1b}b. As evidenced by the graph, it may represent the simpler scheme in our setup where orbits with gain and with losses coexist, thus realizing the basic scenario for chaotic diffusion with gain. This is indeed the case, as shown in Fig.\ref{fig3}. Using a time-resolved analysis of the amplitude behavior of the peaks, the dynamics of the emission intensity at $\nu=163.57$~MHz is reported in the Fig.\ref{fig3}a-b. The emission shows rare, large spikes with strong intermittent
fluctuations both in amplitude and time duration. In Fig.\ref{fig3}c we present the histograms $P(I)$ of such emission, exhibiting a clear evidence of a L\'evy process (magenta dashed line). For the same pump current, other peaks display instead a log-normal distribution; we shown e.g. in the figure the case of $\nu=191.94$~MHz (yellow). The same results can be found in the configuration in Fig.\ref{fig1b}c (a distribution is shown in the inset of Fig.\ref{fig3}c) and, in spite of the different location of the gain (now in the link (4)), in the configuration in Fig.\ref{fig1b}d.

The fact that anomalous fluctuations only occur at some beating frequencies can be justified as follows. In analogy with the random laser case \cite{Lepri2007}, 
we expect fluctuations to occur only for modes which are close enough to threshold, 
namely those with $|\mu_n|<\varepsilon$ with $\varepsilon$ being some small characteristic value. If, as suggested by Fig.\ref{fig2}c, the poles density is roughly uniform, the typical spectral separation between such modes is inversely proportional to  $\varepsilon$. Thus, strongly fluctuating modes will only 
contribute to the subset of the beatings lying at a distance of order 
$1/\varepsilon$, while only far-from-threshold modes will contribute 
to the remaining, yielding a log-normal statistics.

To further check the interpretation, we consider the two cases of Fig.\ref{fig1b}a and Fig.\ref{fig1b}e. In the former, only orbits with gain are possible, thus indicating the settling of saturated, log-normally distributed peaks at the cavity resonances. This is indeed expected as the configuration realizes the standard, mono-directional ring laser: without the isolator the network would be represented by two {\it disjoint} active graphs, with a bistable statistics describing the random jumps between the two emitted modes. 

In the latter, a more complicate situation appears due to the effect of two gains. We first consider that the link (1) only in Fig.\ref{fig1b}e provides gain (the link (4) is now passive, i.e. its laser pump current is kept sufficiently low). Here, we found the same phenomenology reported above; an example of the L\'evy statistics for a peak is reported in Fig.\ref{fig3}d. 
As a second case, we increase the pump current in link (4) so that it has gain as well. The distribution now shows a multistable shape (see inset); an inspection of the equivalent graph reveals that {\it there aren't orbits without gain}. Thus, besides the complex geometry of the network we are dealing with an extension of the ring laser: the multistability arises from the system switching between the saturated modes of the cavity. 

To strengthen the theoretical description, we have compared the experimental findings with a Monte-Carlo type of model (see Suppl. Material). The simulation scheme propagates a set of rays through the network assigning an intensity each that grows or decreases depending on the local value of the gains $g_j$. Whenever a ray reaches a splitter, it is transmitted towards one of the connected fiber chosen at random: the transition probability is given by the splitting factors. Altogether, the ray motion is a random walk on the graph, while the accumulated intensity depends on the whole walk history. The simulation data (for the case of Fig.\ref{fig1b}c) is presented in Fig.\ref{fig3}e, showing the temporal behavior of the intensity collected at one point of the network. The distribution of the data is well fitted with a L\'evy distribution (Fig.\ref{fig3}f).

To conclude, we have presented the LANER as an optical scheme showing laser action and characterized by a fully controllable topological disorder. We studied its emission statistics at the laser threshold, evidencing heavy-tails fluctuations with L\' evy distribution. They are the typical signature of the interplay of chaotic diffusion and amplification of the photons in the network and indicate that the LANER represents an intermediate case between the standard and the random laser.
In particular, the possibility to include multiple and independent gain sections is shared with the latter, but the scheme we have introduced has several advantages in terms of flexibility and control. For instance, it is continuous-wave pumped at variance with standard random laser experiments. From the point of view of basic research, the LANER would allow different investigations, ranging from dynamics on networks of increasing complexity to the effect of the cavity topology on the laser emission. 

\begin{acknowledgments}
We thank F. Cherubini, G.P. Puccioni and acknowledge a 
partial support from EU Project No. 289146 (NETT)

\end{acknowledgments}

\newpage
\section{Supplemental material}
\section{Experimental details}

\begin{figure*}
\includegraphics[width=0.9\linewidth]{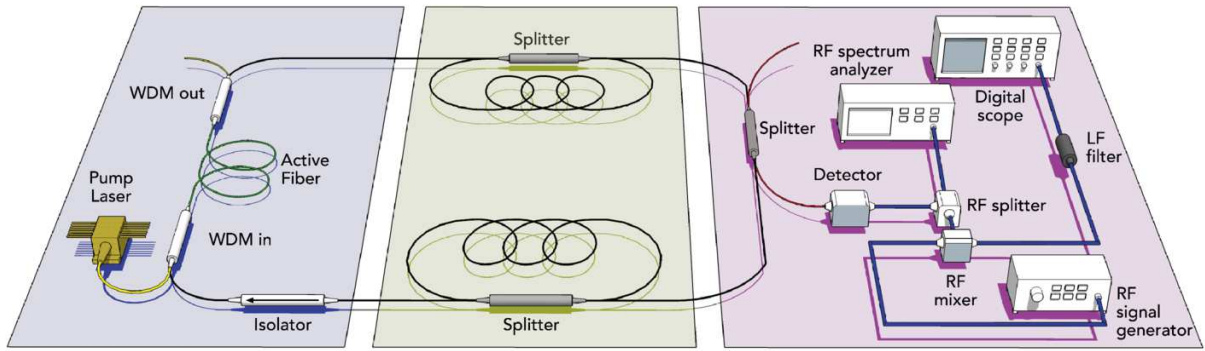}
\caption{(Color online) Setup for the case of Fig.1c of the main text. Left, (oriented) active block composed by a $980~nm$ pumped, $Er^{3+}$ fiber, $WDM$s combining and separating pump and $1.55~\mu m$ signals and an optical isolator. This scheme is represented  as arrowed (red) segments in Fig.1 of the main text.  Center, the rest of the network. Right, the frequency and time-domain detection/acquisition section. }
\label{fig1}
\end{figure*}

An example of the experimental configuration is presented in Fig.\ref{fig1}, 
corresponding to the scheme (c) of Fig.1 of the main text.

The gain in the active links is provided by 1.5~m Erbium-doped optical fiber(s), pumped by $980~nm$ power laser(s). Other optical amplification methods can be used as well such as SOAs etc.
In the setup of Fig.\ref{fig1}, the gain section (pictured in the left side) is composed also by input and output wavelength demuxs (WDMs) to couple and decouple the pump beam to/from the $1550~nm$ signal beam. An optical isolator assures the unidirectional propagation of the signal beam in the gain link. The currents of the pump lasers represent the main control parameters of our setup, as they are directly related to the field amplification in the active links: an arbitrary number of them can be active simultaneously and independently. Throughout the text we will denote the pump current by $J$ whenever a single active fiber is present and with $J_j$ in the case of two (or more). The index $j$ indicates the number of the corresponding link (see the caption of Fig.3 of the main text).

The network section (central part of Fig.\ref{fig1}) is build using single-mode optical fiber and standard, ${\it 2\times2}$ loss-less optical couplers. A ${50:50}$ splitting ratio has been used for all the network splitters in the configurations here considered. Different ratios or number of ports may be used; besides, alternative standard fiber communication devices such as circulators can also be used to implement other coupling configurations. 

The detection is performed by inserting in a network link a 
${50:50, \it 2\times2}$ optical coupler. The split field intensity is collected in the proper direction either with a high bandwidth ($8GHz$) or low bandwidth ($150MHz$), high sensitivity PIN photodetectors. The detector current is then sent to a RF spectrum analyzer and/or a fast digital oscilloscope. 

The measurement of the intensity histograms of a certain frequency peak is carried out acquiring a small frequency interval with the RF spectrum analyzer around the target frequency and storing the maximum value for every scan; the measurement parameters were checked to insure a consistent procedure.  

The time-resolved, amplitude dynamics of a given frequency is obtained by beating the detected signal with an oscillator reference waveform at such frequency, using a RF mixer and then filtering out the high frequency components of the beating signal. The output is then acquired with a digital scope. 
The detection/measuring section is depicted in the right part of Fig.\ref{fig1}.

\section{Calculation of the optical spectra} 

The calculation of stationary modes of 
the network is based on standard linear optics
and conceptually similar to the case of the computation of spectrum of
a quantum graphs \citep{kottos1999periodic,kottos2003quantum}.
We look for solutions where the envelope of the optical field on each
fiber is given by two counter-propagating plane waves. 

For definiteness, let us consider first a single 2$\times$2 loss-less coupler.
We define, at each of the four ports, the column vectors of input and output amplitudes 
$A = (A_1,A_2,A_3,A_4)^T$ and  $B = (B_1,B_2,B_3,B_4)^T$ (this requires fixing 
a port numbering). They are related via the a scattering matrix $B = S^{(1)}A$.
The splitting of the field follows an unitary transformation (energy conservation through the splitter). Under a prescribed convention, we can parametrize it through 
the transmission and reflection coefficients $T,R$ such that $|T|=\sqrt{\alpha}$ and  
$|R|=\sqrt{1-\alpha}$ where $\alpha$ is also termed 
as the \textit{splitting factor} ($\alpha=1/2$ throughout the paper). 
Moreover, the splitters are reflectionless and this 
impose some further  constraint on the form $S^{(1)}$. Altogether
\begin{equation}
S^{(1)} = \left( \begin{array}{cccc}
0 & 0 & \sqrt{1-\alpha} & \sqrt{\alpha} \\
0 & 0 & -\sqrt{\alpha} & \sqrt{1-\alpha} \\
\sqrt{1-\alpha} & -\sqrt{\alpha} & 0 & 0 \\
\sqrt{\alpha} & \sqrt{1-\alpha} & 0 & 0 \end{array} \right) 
\label{S}
\end{equation}
The blocks connecting $\{A_1,A_2\}$ with $\{B_3,B_4\}$ and $\{A_3,A_4\}$ with $\{B_1,B_2\}$ (upper right and bottom left respectively) are unitary matrix as noted before. 
As a consequence, the matrix $S^{(1)}$ is unitary as well.

In the case of $N_s$ splitters that, for simplicity, we assume to be all equal as above, 
we can extend the definitions introducing  
the $4N_s$-dimensional column vectors of input and output amplitudes 
$A$ and $B$ whose components are taken ordered in such a way that 
\begin{equation}
B = S A
\label{Smat}
\end{equation}
and the complete scattering matrix is in block-diagonal form 
\[ S  =
 \begin{pmatrix}
  S^{(1)} & 0 & \cdots & 0 \\
  0 & S^{(1)} & \cdots & 0 \\
  \vdots  & \vdots  & \ddots & \vdots  \\
  0 & 0 & \cdots & S^{(1)}
 \end{pmatrix} ~. \]
The unitary matrix $S$ describes the transfer properties of the splitters (or, in general, of the optical components at the nodes of the network) and it is not dependent from the gains 
and from the topology of the network. It is readily generalized to the case
of a set of different splitters each having different transmission coefficients and/or 
port numbers.

Within a linear description of the network, we assume that the field though the link $j$ (of length $L_j$) is multiplied by a factor $G_j = g_j e^{i K L_j}$, 
According to the definition of the field variables, we can write this condition 
in terms of the {\it propagation matrix} $P$
\begin{equation}
A = P B~,
\label{Pmat}
\end{equation}
where in a fully connected network $N_l = 2 N_s$; $P$ contains the full information on 
the topology of the network via the \textit{connectivity (adjacency) matrix} 
(how the links are connected through the splitters) 
and on the gain in the links. It has some properties dictated by the 
physical nature of the connections: (i) since every port of all splitters is connected to one and only one (different) port, $P$ has exactly one element different from zero 
in every row and in every column; (ii) since a port cannot be connected to itself, 
the diagonal elements of $P$ vanish; (iii) $\det P = |G_1|^2|G_2|^2\ldots |G_{N_l}|^2$.

Combining equations (\ref{Pmat}) and (\ref{Smat}) and imposing
that the resulting linear homogeneous systems has nontrivial solutions yields 
the condition (1) given in the main text with $N=P(K)S$.

We also checked that the integrated density of states, namely the number of modes below $k$ grows on average linearly, 
as expected from the Weyl law for one-dimensional structure
\citep{kottos1999periodic}.

The above definitions require some minor adaptation to account for the presence of optical isolators as in this case some components are not allowed to propagate
and thus the corresponding elements of $P$ are zero. 

A more complete description of the mathematical details will be given elsewhere.

\section{Mapping the network on graphs}

The optical network is mapped on a directed graph using the following procedure. A new matrix $M$ is built from $N$ by replacing its non-zero elements with $1$'s; $M$ is then interpreted as the adjacency matrix of the associated graph. Each vertex of the graph represents the optical mode propagating along one of the two allowed directions in a link; conventionally, we plot it as a empty-red (filled-black) dot whether such link does (does not) contain a gain element. When a link accommodates an optical isolator, the vertex in the graph corresponding to the blocked mode is removed (the related element in the adjacency matrix is zero). As a consequence, the graph is {\it pruned}: the above vertex and all the incoming/outcoming bonds connected to it are erased. We remark that, the lasing action can happen only if at least a empty-red vertex is present in the graph., i.e. networks whose graphs contain only filled-black vertices cannot lase and therefore be detectable. 

\section{Beating and length spectra}
 
The observable is the field in the links, detected inserting additional power splitters; both propagation directions are accessible. The signal is detected with photodetectors, yielding an electric signal proportional to the intensity of the field. As a consequence,  in the present setup, we cannot directly access to the full information of the field, but only to the 
\textit{beating of all the optical modes} within the observational bandwidth (see the panels in Fig. 2)
To compare the measurements with the theory, we thus start from the optical spectrum numerically computed as described and we proceed as follows. 
From a given set of $M$ values of the wavenumbers $k_n$ we 
compute all the differences $k_n-k_m$ belonging to a given interval $(0,B)$, with $B$
being an assigned bandwidth. An histogram of such data is than evaluated, comparing to the experimental RF spectra. The numerical results and the measurements are reported in the 
Figs.2d-e.

Another measured quantity is 
\begin{equation}
F(l) = |\sum_n w(\nu_n) \exp(i\nu_n l)|^2
\label{lensp}
\end{equation}
where $\nu_n$ are the experimentally measured positions of the peaks (evaluated by a quadratic interpolation around the relative maxima) and $w$ is a window function $w(x)=\sin^2(\pi x/x_{max})$ where $x_{max}$ is the largest value in the set.
It can be shown that this observable is approximately proportional to the so-called length spectrum, namely the squared modulus  of the Fourier transform of the 
density of states. The latter has been used to characterize the 
spectra of quantum graphs in terms of the classical periodic orbits 
\citep{kottos2003quantum}. In the numerical calculations, where we have access
to the actual wavenumbers $k_n$ we checked that definition (\ref{lensp}) yields
the same peak structure as the length spectrum itself.

\section{Monte-Carlo simulations}
 
The evolution is simulated in terms of ray dynamics. A discrete-time evolution scheme is implemented over the network of interest with a Monte-Carlo approach. In a single realization, a starting initial condition is prepared as a ray in a link; at every time step, it moves in the link along the initial direction. When a splitter is approached, the ray chooses one of the two possible exits selected according to the outcome of a continuous, uniform random process with probability 0.5 (in general, with the branch-specific splitting ratio of the splitter used) and it is amplified according to the former link gain (or attenuation), with a saturation nonlinearity that prevents the dynamics to diverge. The detection is carried out by reading the intensity at every step in a fixed position on a link (eulerian detection). To improve the statistics, an initial ensemble of rays has been used, corresponding to fully populate a link with pulses and let them evolve.

\end{document}